# On Generalized Tian Ji's Horse Racing Strategy

Jian-Jun SHU
School of Mechanical & Aerospace Engineering, Nanyang Technological University,
50 Nanyang Avenue, Singapore 639798

**ABSTRACT**: Tian Ji's horse racing strategy, a famous Chinese legend, constitutes a promising concept to be applied to important issues in today's competitive environment; this strategy is elaborated on and analyzed by examining the general case. The mathematical formulation concerning the calculation of winning, drawing or losing combinations and probabilities is presented to illustrate the interesting insights on how ancient philosophies could promote thinking in business competitiveness, in particular, the wisdom behind *sacrificing the part for the benefit of the whole* or *sacrificing the short-term objectives in order to gain the long-term goal*.

## 1 Introduction

In the modern world today, the survival of the fittest holds for general people, but the survival of the cunning is possible for intellectuals. Nowadays, with the development of society, most companies face more intense competition, especially for the weaker and smaller ones. In actuality, the stronger and larger companies possess many more advantages and better talents than the weaker and smaller ones. Despite the fact, the strongest of warriors has his Achilles' heel. In such a competitive world whereby almost everything can be rivaled in, it is necessary for the weaker and smaller companies to plan the order of engagement in order to compete with stronger and larger rivals. This is just like playing a team sport. The coach of the weaker team must carefully organize his players in suitable order of appearance. The idea is *to sacrifice the part for the benefit of the whole* or *to sacrifice the short-term objectives in order to gain the long-term goal*. This procedure is very similar to Tian Ji's horse racing strategy, a famous Chinese legend. In this study, the generalized Tian Ji's horse racing strategy is analyzed mathematically to provide an interesting insight into decision making in dynamic and highly competitive global environments. The correct application of the principle of gaining overall victory with partial loss may enable the inferior side to gain superiority and win victory with a surprise move.

## 2 Tian Ji's horse racing strategy

In ancient China, there was an era known as "Warring States Period" (403 BC — 221 BC) during which China was not a unified empire but divided by independent Seven Warring States with conflicting interests, one of which was Qi State located in eastern China. From 356 BC to 320 BC, the ruler of Qi State was Tian Yin-Qi (378 BC — 320 BC), King Wei of Qi. The story of Tian Ji's horse racing strategy, which is well-known and popular in China today, was originally recorded [1] in the biography of Sun Bin (? — 316 BC), as a military strategist in Qi State ruled by King Wei of Qi:

> *General Tian Ji, a high-ranking army commander in Qi State, frequently bet heavily on horse races with King Wei of Qi. Observing that their horses, divided into three different speed classes, were well-matched, Sun Bin then advised Tian Ji, "Go ahead and stake heavily! I shall see that you win." Taking Sun Bin at his word, Tian Ji bet a thousand gold pieces with the King. Just as the race was to start, Sun Bin counseled Tian Ji, "Pit your slow horse against the King's fast horse, your fast horse against the King's medium horse, and your medium horse against the King's slow horse." When all three horse races were finished, although Tian Ji lost the first race, his horses prevailed in the next two, in the end getting a thousand gold pieces from the King.*

Amazedly, the victorious strategy (as did Tian Ji after following Sun Bin's advice) was remarkable to be achieved 2300 years long before operations research and game theory were invented [2]. This was only one way that Tian Ji could claim a victory over the King, as illustrated in Figure 1. All the other options would present Tian Ji with loss. Sun Bin's victorious advice, since called Tian Ji's horse racing strategy, can be extended to a scenario where Tian Ji and the King would race horses within a disjoint stratification of an arbitrary number $N$ different speed classes. In order to facilitate the analysis of the generalized Tian Ji's $N$-horse racing strategy, the $N$ horses owned by two players: Tian Ji ($T$) and the King ($K$) are denoted respectively by $T_n$ and $K_n$, where the subscript $n = 1, 2, \cdots, N$ is defined as player $T$'s or $K$'s horse in the $n$th speed class. In this scenario of $N$-horse racing, $T$'s horse in the faster class is able to beat $K$'s horse in the slower class, but $T$'s horse is unable to beat $K$'s horse in the same or faster class. Without losing generality, the relative racing capabilities of horses are $T_{n+1} \prec K_{n+1} \prec T_n \prec K_n$ for any $n = 1, 2, \cdots, N-1$, where the symbol "$\prec$" means "unable to beat" and the larger subscript corresponds to the slower class.

Figure 1: Tian Ji's horse racing strategy

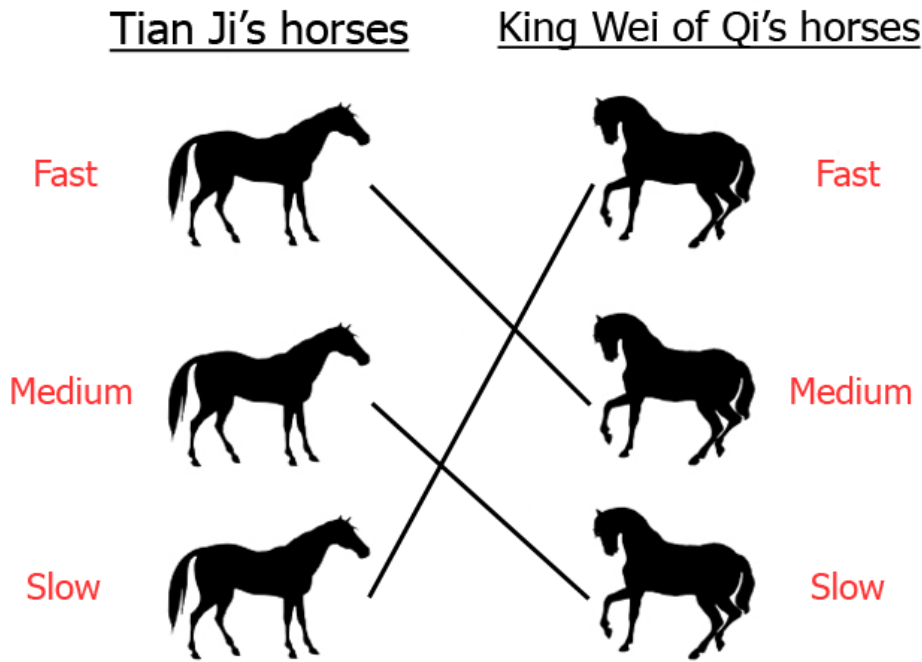

$T$ and $K$ would choose the same class of $N$-horse racing, that is, the pairwise racing is $\begin{pmatrix} T_1 & T_2 & \cdots & T_N \\ K_1 & K_2 & \cdots & K_N \end{pmatrix}$. Because $T$'s horse is slower than $K$'s one in the same class $(T_n \prec K_n,\ n = 1,2,\cdots,N)$, $T$'s horses would lose all races. The essence of Tian Ji's horse racing strategy is that the originally-classified racing appearance of $T$'s horses should be shifted one place in order to achieve the $T$'s best result. The best strategy for the generalized Tian Ji's $N$-horse racing, suggested by Sun Bin, should be the pairwise racing $\begin{pmatrix} T_N & T_1 & \cdots & T_{N-1} \\ K_1 & K_2 & \cdots & K_N \end{pmatrix}$. $T$ would claim a victory of the $N$-horse racing with one loss and $N-1$ wins.

If $K$ always chooses the racing appearance $(K_1, K_2, \cdots, K_N)$ against $T$'s best response $(T_N, T_1, \cdots, T_{N-1})$, $T$ would win every bet. Naturally, $K$ would soon realize that the racing appearance $(K_1, K_2, \cdots, K_N)$ is resulting in recurrent losses. $K$ would become an active player and consider an alternative racing appearance to turn the racing around. A competitive situation is encountered for each player competing with a total of $N!$ combinatorial pairwise racing $\begin{pmatrix} T_{\sigma(1)} & T_{\sigma(2)} & \cdots & T_{\sigma(N)} \\ K_1 & K_2 & \cdots & K_N \end{pmatrix}$ (where $\sigma$ is the permutation) available to $T$ and $K$. The above explanation indicates that $T$ would lose all races for the unit permutation $\sigma(n) = n$ (as $T$'s worst strategy) and $T$ would claim a victory with one loss and $N-1$ wins for the shift permutation $\sigma(n) = \begin{cases} N & n = 1 \\ n-1 & 1 < n \leq N \end{cases}$ (as $T$'s best strategy). Then the natural question is what is $T$'s winning probability for randomly-pairwise racing between $T$'s and $K$'s horses. The equivalent question is how many permutations are available to $T$ as $T$'s victorious strategies.

**Theorem 1**: The number of $T$ having exactly $M$ wins in $N$-horse racing is the Eulerian number [3], $E(N,M) = \sum_{m=0}^{M} \left[ (-1)^m (M-m+1)^N \frac{(N+1)!}{m!(N-m+1)!} \right]$, the number of permutations on $\{1,2,\cdots,N\}$ with exactly $M$ excedances.

*Proof* An excedance of the permutation $\sigma$ on $\{1,2,\cdots,N\}$ is defined as any index $n$ such that, $\sigma(n) > n$ and the Eulerian number $E(N,M)$ is defined as the number of the permutation $\sigma$ on $\{1,2,\cdots,N\}$ with exactly $M$ excedances [4]. It is obvious that the existence condition of $T$ having one win is $K_n \prec T_{\sigma(n)}$, that is, $\sigma(n) > n$ for any index $n$. To determine the number of $T$'s wins is equivalent to counting the number of $\sigma(n) > n$, which is an excedance of the permutation $\sigma$ in the parlance of combinatorics. So the number of $T$ having exactly $M$ wins in $N$-horse racing is the Eulerian number $E(N,M)$. The detailed proof for the summation formula of the Eulerian number, $E(N,M) = \sum_{m=0}^{M} \left[ (-1)^m (M-m+1)^N \frac{(N+1)!}{m!(N-m+1)!} \right]$, is given as follows.

There are two ways of getting an $N$-permutation with $M$ excedances from an $(N-1)$-permutation by inserting the entry $N$. Either the $(N-1)$-permutation has $M$ excedances, and the insertion of $N$ does not form a new excedance, or $M-1$ excedances, and the insertion of $N$ does form a new excedance.

In the first case, the entry $N$ is placed at the end, or at the position of one of the $M$ excedances and the replaced one is moved into the end. In the second case, the entry $N$ is placed at the position of one of the $(N-1)-(M-1) = N-M$ non-excedances and the replaced one is moved into the end. The desired recurrence is obtained as

$$E(N,M) = (M+1)E(N-1,M) + (N-M)E(N-1,M-1) \quad \text{for all } N \geq 2. \tag{T1}$$

Note that for binomial coefficient, $\binom{N}{M} = \frac{N!}{M!(N-M)!}$,

$$n\binom{M+n}{N-1} = (M+1)\binom{M+n}{N} + (N-M-1)\binom{M+n+1}{N}.$$

Therefore, we have

$$n \sum_{M=0}^{N-2} \left[ E(N-1,M) \binom{M+n}{N-1} \right]$$

$$= \sum_{M=0}^{N-2} \left\{ E(N-1,M) \left[ (M+1)\binom{M+n}{N} + (N-M-1)\binom{M+n+1}{N} \right] \right\}$$

$$= \sum_{M=0}^{N-2} \left[ (M+1)E(N-1,M)\binom{M+n}{N} \right] + \sum_{M=1}^{N-1} \left[ (N-M)E(N-1,M-1)\binom{M+n}{N} \right]$$

$$= \sum_{M=0}^{N-1} \left[ E(N,M)\binom{M+n}{N} \right],$$

where the last step uses (T1) and $E(N,-1) = E(N,N) = 0$. By mathematical induction on $N$, the above expression can be used to prove the Worpitzky's identity

$$n^N = \sum_{M=0}^{N-1} \left[ E(N,M)\binom{M+n}{N} \right] \qquad \text{for all } N \geq 1. \tag{T2}$$

Using (T2) with $\binom{n}{m} = \binom{n}{n-m}$, $\binom{n}{m} = (-1)^m \binom{-n+m-1}{m}$, $\binom{n_1+n_2}{n} = \sum_{m=0}^{n} \binom{n_1}{m}\binom{n_2}{n-m}$ and $\binom{0}{m} = \begin{cases} 1 & m=0 \\ 0 & m \neq 0 \end{cases}$, we get

$$\sum_{m=0}^{M} \left[ (-1)^m (M-m+1)^N \frac{(N+1)!}{m!(N-m+1)!} \right] = \sum_{m=0}^{M} \left[ (-1)^m (M-m+1)^N \binom{N+1}{m} \right]$$

$$= \sum_{n=0}^{M} \left\{ E(N,n) \sum_{m=0}^{M-n} \left[ (-1)^m \binom{N+M-n-m}{N}\binom{N+1}{m} \right] \right\}$$

$$= \sum_{n=0}^{M} \left\{ E(N,n) \sum_{m=0}^{M-n} \left[ (-1)^m \binom{N+M-n-m}{M-n-m}\binom{N+1}{m} \right] \right\}$$

$$= \sum_{n=0}^{M} \left\{ E(N,n)(-1)^{M-n} \sum_{m=0}^{M-n} \left[ \binom{-N-1}{M-n-m}\binom{N+1}{m} \right] \right\}$$

$$= \sum_{n=0}^{M} \left[ E(N,n)(-1)^{M-n} \binom{0}{M-n} \right] = E(N,M).$$

The theorem is proved. ∎

In view of the symmetry property of the Eulerian number, that is, $E(N,M) = E(N, N-(M+1))$, **Theorem 1** can be expressed equivalently as follows in **Theorem 2**.

**Theorem 2**: The number of $T$ having exactly $M+1$ no-wins in $N$-horse racing is the Eulerian number.

In both theorems, it is interesting to note that the generalized Tian Ji's horse racing strategy, as the extension of the famous Chinese legend, can be viewed as a practical demonstration of applying the Eulerian number. There are only three outcomes for $T$, namely the winning combination $\begin{cases} \sum_{M=\frac{N+1}{2}}^{N-1} E(N,M) & \text{odd } N \\ \frac{N!}{2} - E(N,\frac{N}{2}) & \text{even } N \end{cases}$ with probability $\begin{cases} \frac{1}{N!} \sum_{M=\frac{N+1}{2}}^{N-1} E(N,M) & \text{odd } N \\ \frac{1}{2} - \frac{1}{N!} E(N,\frac{N}{2}) & \text{even } N \end{cases}$, the drawing combination $\begin{cases} 0 & \text{odd } N \\ E(N,\frac{N}{2}) & \text{even } N \end{cases}$ with probability $\begin{cases} 0 & \text{odd } N \\ \frac{1}{N!} E(N,\frac{N}{2}) & \text{even } N \end{cases}$, or the losing combination $\begin{cases} \sum_{M=0}^{\frac{N-1}{2}} E(N,M) & \text{odd } N \\ \frac{N!}{2} & \text{even } N \end{cases}$ with probability $\begin{cases} \frac{1}{N!} \sum_{M=0}^{\frac{N-1}{2}} E(N,M) & \text{odd } N \\ \frac{1}{2} & \text{even } N \end{cases}$, which are shown in Table 1.

Table 1: Combination and probability with variable number of horses

| $N$ horses | Total $N!$ | Winning | Drawing | Losing |
|---|---|---|---|---|
| | | Combination (with probability) | | |
| 1 | 1 | 0 (0%) | 0 (0%) | 1 (100%) |
| 2 | 2 | 0 (0%) | 1 (50%) | 1 (50%) |
| 3 | 6 | 1 (17%) | 0 (0%) | 5 (83%) |
| 4 | 24 | 1 (4%) | 11 (46%) | 12 (50%) |
| 5 | 120 | 27 (23%) | 0 (0%) | 93 (78%) |
| 6 | 720 | 58 (8%) | 302 (42%) | 360 (50%) |
| 7 | 5040 | 1312 (26%) | 0 (0%) | 3728 (74%) |
| 8 | 40320 | 4541 (11%) | 15619 (39%) | 20160 (50%) |
| 9 | 362880 | 103345 (29%) | 0 (0%) | 259535 (72%) |
| 10 | 3628800 | 504046 (14%) | 1310354 (36%) | 1814400 (50%) |

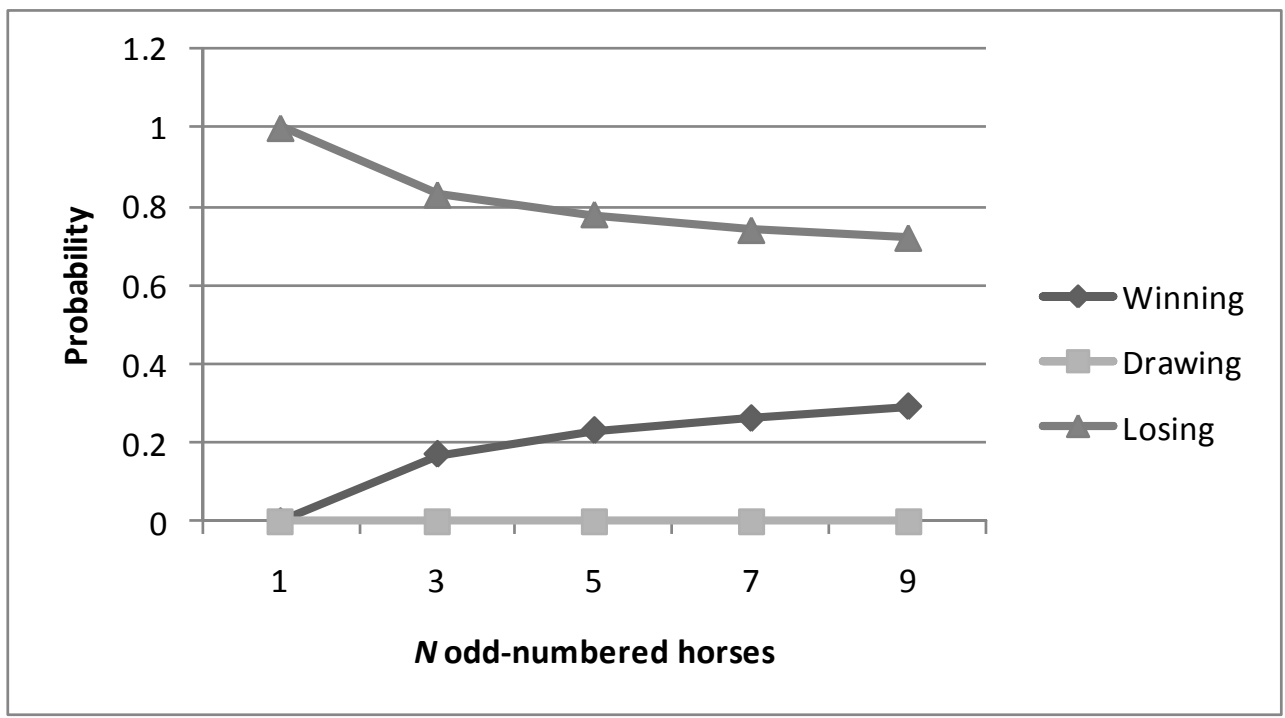

Figure 2: Trend of probability for odd-numbered horses

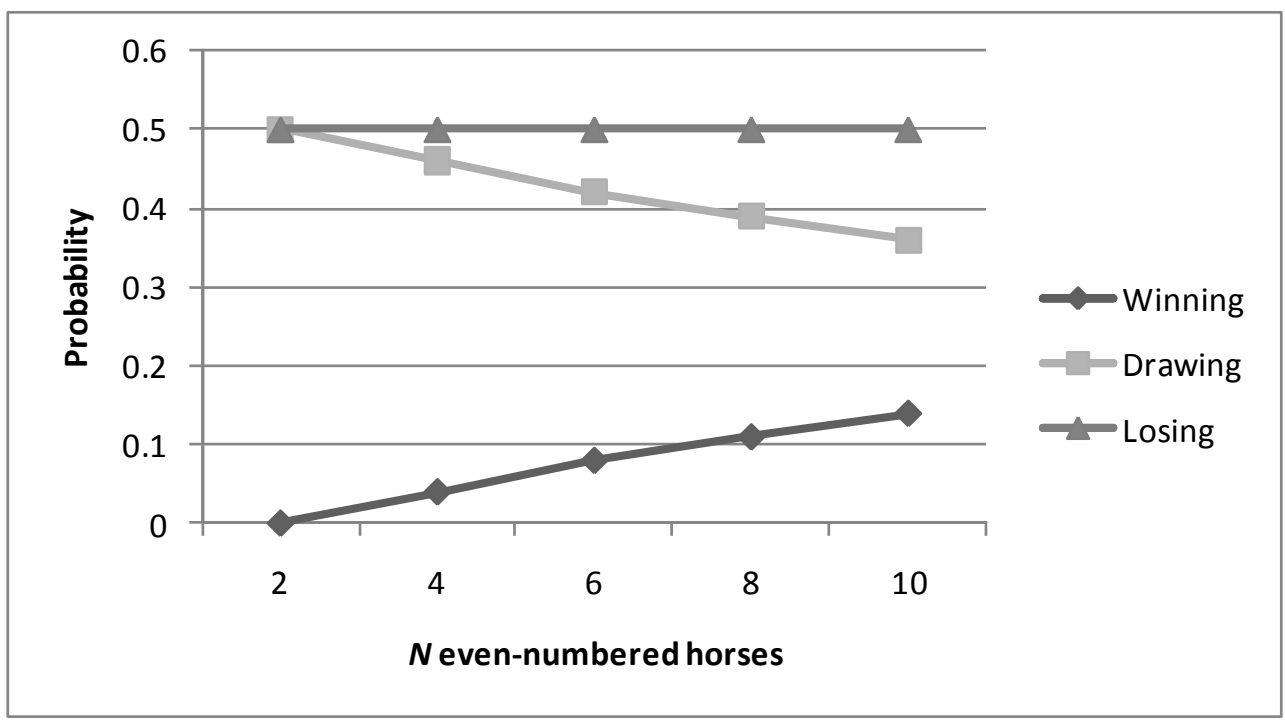

Figure 3: Trend of probability for even-numbered horses

Probabilities for odd- or even- numbered horses are plotted respectively in Figures 2 and 3. From the results illustrated above, the probabilities follow the Eulerian distribution and there are two detectable characteristics. First, the case of odd-numbered horse racing has no drawing, which drawing happens only in the case of even-numbered horse racing. Second, the losing probabilities of any even-numbered horse racing are always at the constant 50% regardless of horse number involved.

In the odd-numbered horse racing, the winning and losing probabilities converge to the constant 50% as horse number increases due to no drawing; whereas in the even-numbered horse racing, the winning and drawing probabilities converge to a constant 25% as horse number increases due to the constant 50% losing. Overall, the winning probability of the odd-numbered horse racing is much higher than that of the adjacent even-numbered cases, for example, 23% of $N = 5$ is much higher than 4% of $N = 4$ and 8% of $N = 6$.

This shows that odd-numbered horse racing gives an opportunity of winning better than an even-numbered case does. This implies that the best combat units should be odd-numbered. Of course, as the combat efficiency would be getting better as $N$ increases, the difficulty of controlling much large $N$ combat units would be encountered. No drawing occurs in the odd-numbered horse racing, which means that a decisive outcome must be reached instead of a stalemate. More importantly, Figures 2 and 3 suggest that the more horses involved, the larger $N$, the higher is the winning probability. Philosophically, it is typically the epitome of winning in numbers.

## 3 Concluding Remarks

This paper is a generalization of the mathematical version of Tian Ji's horse racing strategy involving a one-to-one contest between two sets of racing horses within a disjoint stratification of speed classes. The formulation of determining the winning, drawing and losing probabilities of the generalized Tian Ji's horse racing strategy for any given number of racing horses is discussed. Based on the Eulerian number, the way of calculating the number of having exactly $M$ wins in $N$-horse racing is straightforward, thereby enabling us to find the probability of winning an entire game by having more wins than losses. The wisdom behind Tian Ji's horse racing strategy is *to sacrifice the part for the benefit of the whole* or *to sacrifice the short-term objectives in order to gain the long-term goal*. As an example of the generalized Tian Ji's horse racing strategy, the mathematical treatment of how to sacrifice could promote philosophical thinking in dealing with complicated situations [5]. In business, product diversification requires very different financial, human and technological resources. Trade-offs are inevitable. A company has to decide which part should be sacrificed for the benefit of the whole or which short-term objectives should be sacrificed in order to gain the long-term goal. Studying the theory of Tian Ji's horse racing strategy provides interesting insight into decision making in dynamic and highly competitive global environments.